\documentclass[12pt]{iopart}

\usepackage{graphicx}% Include figure files
\usepackage{epsfig}
\usepackage{hhline}
\usepackage{iopams}  %   ,amssymb,amsfonts}

\usepackage[usenames]{color}

\begin{document}
\title[Running Gluon Mass ...]{Running Gluon Mass from Landau Gauge Lattice QCD Propagator}
\author{O. Oliveira}
\address{Departamento de F\'{\i}sica, Universidade de Coimbra, 3004-516 Coimbra, Portugal \\
                \mbox{ } \\
                Departamento de F\'{\i}sica, Instituto Tecnol\'ogico de Aeron\'autica, 
                               12.228-900 S\~ao Jos\'e dos Campos, SP, Brazil }
\ead{orlando@teor.fis.uc.pt}

\author{P. Bicudo}
\address{Departamento de F\'{\i}sica, I.S.T., Av Rovisco Pais, 1049-001 Lisboa, Portugal} 
\ead{bicudo@ist.utl.pt}
\date{\today}
\begin{abstract}
The interpretation of the Landau gauge lattice gluon propagator as a massive type bosonic propagator is investigated. 
Three different scenarios are discussed: i) an infrared constant gluon mass; ii) an ultraviolet constant gluon mass; 
iii) a momentum dependent mass.
We find that the infrared data can be associated with a massive propagator up to momenta $\sim 500$ MeV,
with a constant gluon mass of 723(11) MeV, if one excludes the zero momentum gluon propagator 
from the analysis, or 648(7) MeV, if the zero momentum gluon propagator is included in the data sets. 
The ultraviolet lattice data is not compatible with a massive type propagator with a constant mass.
The scenario of a momentum dependent gluon mass gives a decreasing mass with the momentum, which
vanishes in the deep ultraviolet region. Furthermore, we show that the functional forms used to describe
the decoupling like solution of the Dyson-Schwinger equations are compatible with the lattice data with
similar mass scales. 
\end{abstract}
\pacs{14.70.Dj; 11.15.Ha; 12.38.-t}

\maketitle

%====================================================================
%====================================================================
\section{Introduction and Motivation}

The lagrangian for pure SU(3) Yang-Mills theory does not include a mass scale.  
At the classical level, conformal invariance of pure gauge theories is the expression of this lack of scale. 
However, the corresponding quantum theory gets a mass, let us say $\Lambda_{QCD}$, from the 
loop contributions via dimensional transmutation. 

At the level of the lagrangian a gluon mass term is forbidden by gauge invariance and, as long as the color 
symmetry is unbroken, the gluon is supposed to be massless. 
Certainly, in what concerns the perturbative solution of QCD, within the framework
of the Faddeev-Popov quantization procedure \cite{FaPo67}, the gluon is massless. 
However, as discussed in \cite{Cornwall82}, a dynamical generated mass which is a function of the momentum
is allowed if one goes beyond perturbation theory.

From the theoretical point of view, a non-vanishing gluon mass is welcome to regularize infrared divergences and
to solve some problems related with unitarity. 
On the other side, diffractive phenomena \cite{Forshaw99} and inclusive 
radiative decays of $J / \psi$ and $\Upsilon$ \cite{Field02} suggest a massive gluon. 
Moreover, 
lattice simulations suggest an infrared gluon hard mass of $\sim 600$ MeV \cite{Oliveira09} and an ultraviolet mass 
$M_g \sim 1.0$ GeV \cite{Leinweber99,Silva04}. Phenomenology favors a gluon mass between 
$\sim 0.500$ GeV and $\sim 1.2$ GeV depending on how the mass is defined -- see table 15 in \cite{Field02}.
Furthermore, a dynamically generated gluon mass can be related with the presence of the
$\langle A^2 \rangle$ gluon condensate, which is associated with the non-perturbative sector of QCD and whose
role has been investigated by several authors - see, for example, 
\cite{Dudal04,Esole04,Dudal10} and references their in.

The idea of a gluon mass was explored by different authors
\cite{Aguilar03,Aguilar08,Aguilar08b,Aguilar08c,Aguilar10,Cornwall09,Sauli09,Fischer09}.
Starting from the Dyson-Schwinger equations (DSE) for the gluon and ghost propagators, 
after a suitable truncation scheme,  the equations were
solved and the transverse part of the gluon propagator described by a massive type propagator, i.e.
\begin{equation}
   D(q^2) = \frac{Z(q^2)}{q^2 + M^2 (q^2)} \, ,
   \label{Dmassive}
\end{equation}
with a a momentum dependent gluon mass $M^2 (q^2)$, called below running mass gluon,
and a running dressing function $Z(q^2)$.
Typically, the numerical solution for the propagator is fitted to a functional form $M^2  (q^2)$ copied from
the solution discussed in \cite{Cornwall82}. 
The solutions of the DSE give a $M(q^2)$ which takes its largest value at zero momentum, 
where $M(0) \sim 600$ MeV, and vanishes for $q \gg \Lambda_{QCD}$. 
In this way, the usual perturbative propagator is recovered at high momentum. 

The Dyson-Schwinger studies referred to in the above paragraph rely on the Faddeev-Popov quantization method, whose 
validity for investigating non-perturbative effects in QCD is under debate - see, for example, \cite{Dudal11,Dudal08} 
and references therein. 
However, a running gluon mass was also found 
within the studies of the non-perturbative quantization  of Yang-Mills theories. 
In particular, in the framework of the so-called refined Gribov-Zwanziger action \cite{Dudal08}, a tree level propagator was 
computed suggesting a functional form for $M(q^2)$ which is close to the original proposal of Cornwall
 \cite{Cornwall82}.

Besides the solutions of the Dyson-Schwinger equations, lattice simulations also provide support for a
non-vanishing gluon mass, see for example \cite{Oliveira09,Leinweber99,Silva04,Bernard94}. 
The precise value for $M(q^2)$ depends on how the gluon propagator is modeled. 
For example, in lattice QCD or Dyson-Schwinger calculations, 
a running mass is computed fitting the propagator to a given functional form for $M(q^2)$. 
In this work we use lattice QCD simulations to investigate $M^2(q^2)$. Besides checking the compatibility of
the lattice data with theoretical predictions for $M^2(q^2)$, a first attempt is made to compute the running gluon
mass directly from lattice simulations. 

The paper is organized as follows. 
In section \ref{definicoes} we describe the lattice setup, the cuts performed to produce a unique curve for the
gluon propagator and the renormalization procedure. 
The gluon mass is investigated in section \ref{mass_gluao} considering three different scenarios.
Before considering the running mass, we consider constant mass ansatze to fit the propagator for the 
infrared  (section \ref{section_hadr_ir}) and  the ultraviolet momenta (section \ref{section_hadr_uv}).  
We find that the gluon propagator cannot be fully fitted with a constant gluon mass, and thus we proceed 
with fitting the gluon propagator with a running mass gluon and a running gluon dressing function 
(\ref{section_run_m}).
Finally, in section \ref{fim} we resume the results of section \ref{mass_gluao} and
comment on its interpretation.

%====================================================================
%====================================================================
\section{Definitions and Lattice Setup \label{definicoes}}

\begin{figure}[t]
  \vspace{0.1cm}
   \centering
   \includegraphics[scale=0.6]{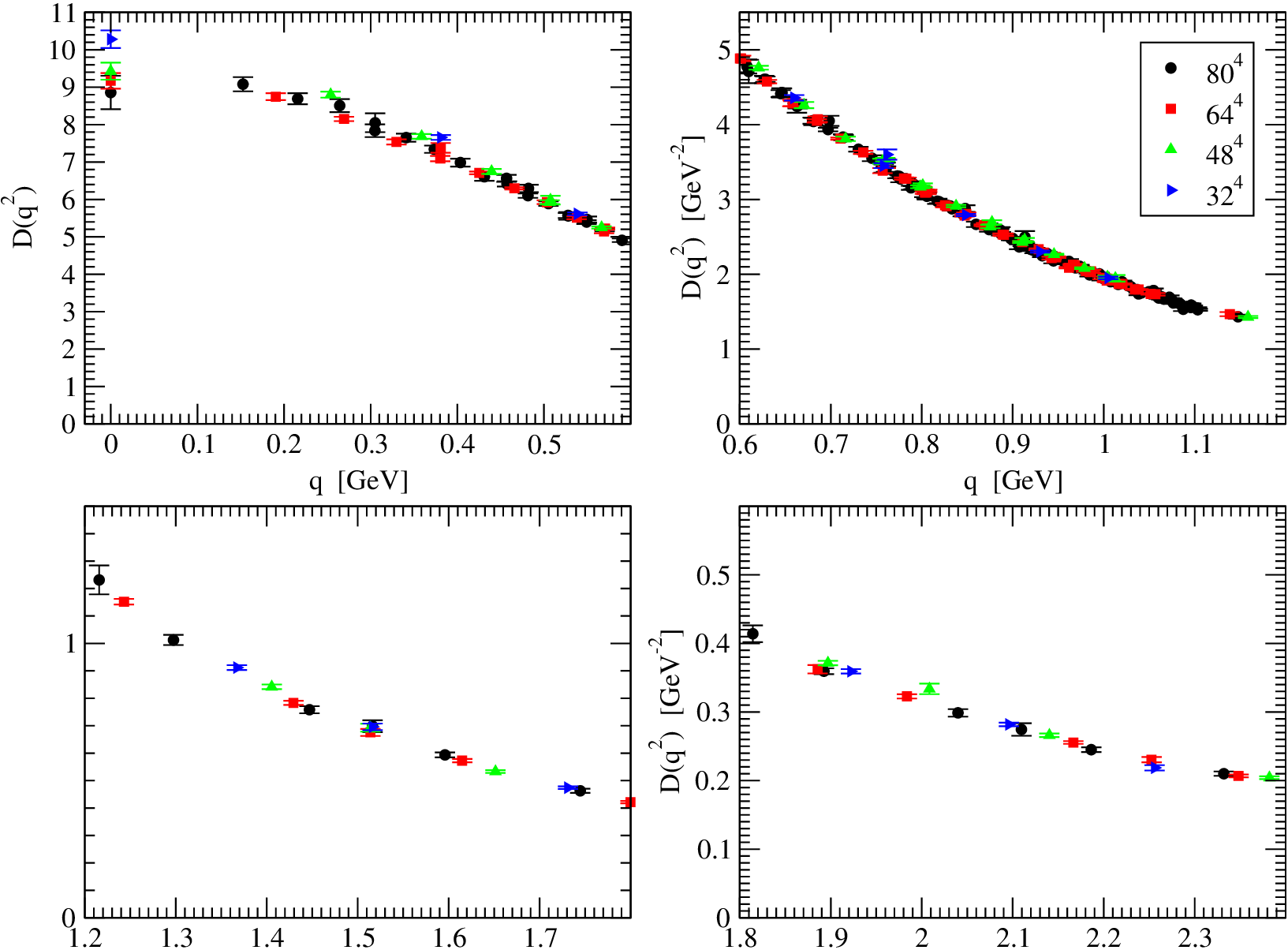} % requires the graphicx package
   \caption{Renormalized gluon propagator for all the lattices described in table \ref{tab_lattices}. 
   The propagator (vertical axis) is given in GeV$^{-2}$ and the momenta (horizontal axis) in GeV.}
   \label{fig_prop_all}
\end{figure}

In this paper we consider four dimensional SU(3) Wilson action lattice simulations at $\beta = 6.0$. For this 
$\beta$ value the lattice spacing, measured from the string tension \cite{Bali93}, is given by 
$a = 0.1016(25)$ fm or, equivalently, $a^{-1} = 1.943$ GeV.

The gauge configurations were generated with the MILC code \cite{milc}. Each configuration, sampled with the 
Wilson action, was rotated to the Landau gauge, as described in \cite{Silva04}, using for gauge fixing the
overrelaxation algorithm. 
The gauge fixing process requires a maximization of a given functional over the gauge orbit of each configuration.
The maximization process was stopped when the lattice tetra-divergence \cite{Silva04}, averaged per site,
become smaller than $10^{-13}$. 
The set of lattices considered in the present work are summarized in table \ref{tab_lattices}.

In the Landau gauge, the gluon propagator is given by
\begin{equation}
   D^{ab}_{\mu\nu} (q^2) = \langle A^a_\mu (q) A^b_\nu \rangle = \delta^{ab} \left(  \delta_{\mu\nu} - \frac{q_\mu q_\nu}{q^2} \right) \, D(q^2) \,;
\end{equation}   
latin letters stand for color indices and greek letters for space-time indices. 
The momentum space gluon field $A^a_\mu (q)$ definition and how to compute the form factor $D(q^2)$ are
described in \cite{Silva04} and will not be repeated here. 

For the continuum momentum we take the standard definition
\begin{equation}
   q_\mu = \frac{2}{a} \, \sin \left( \frac{\pi}{L_\mu} \, n \right) \, , \hspace{0.2cm} n = 0, \dots, L_\mu - 1 \, ,
\end{equation}
where $a$ is the lattice spacing and $L_\mu$ the number of lattice points in direction $\mu$. 

In the following only renormalized data will be considered. 
The renormalization was performed fitting, for each lattice simulation, the bare lattice propagator to the one-loop inspired result
\begin{equation}
   D(q^2) = \frac{K}{q^2} \left( \ln \frac{q^2}{\Lambda^2} \right)^{-\gamma} \, ,
   \label{fit_uv}
\end{equation}   
where $\gamma = 13/22$ is the gluon anomalous dimension. 
For the fits to equation (\ref{fit_uv}), the largest momentum range, starting at $q_{min}$ and going up to $\sim 5$ GeV, with the
$\chi^2/d.of.$ closer to unity was used -- see table \ref{tab_Z}.
From the fits we extracted the constants $K$ and $\Lambda$, which enabled the computation of the renormalization constant $Z_R$ via
\begin{equation}
    D(q^2) = Z_R D_{Lat} (q^2) \, ,
\end{equation}    
after requiring the renormalized propagator to be given by
\begin{equation}
  \left.    D(q^2) \right|_{q^2 = \mu^2} = \frac{1}{\mu^2} \, .
\end{equation}
As renormalization scale it was used $\mu = 3$ GeV.

\begin{table}[t]
   \centering
   \begin{tabular}{l@{\hspace{0.7cm}}r@{\hspace{0.4cm}}r@{\hspace{0.4cm}}r@{\hspace{0.4cm}}r} 
      \hline
 L              &      32      & 48 & 64   & 80 \\
 L(fm)        &      3.2     & 4.9 & 6.5 & 8.1 \\
  \# Confs. &    126    & 104 & 120 & 50 \\
      \hline
   \end{tabular}
   \caption{Lattice setup - the lattices considered are $L^4$ symmetric hypercubes.
                 For conversion into physical units the lattice spacing we used $a = 0.1016(25)$ fm, or $a^{-1} = 1.943$ GeV, as
                 computed from the string tension \cite{Bali93}. }
   \label{tab_lattices}
\end{table}

\begin{table}[t]
   \centering
   \begin{tabular}{l@{\hspace{0.7cm}}r@{\hspace{0.5cm}}r@{\hspace{0.5cm}}r@{\hspace{0.5cm}}r} 
      \hline
 L                                           &  32                & 48              & 64              & 80 \\
 $q_{min} - q_{max}$ (GeV)  &  2.81 - 5.08  & 2.49 - 5.02 & 1.51 - 5.14 & 1.52 - 5.05 \\
 $\chi^2/d.of.$                        &  0.91            & 0.97           & 0.89            & 1.06 \\
 $Z_R$ (GeV$^-{2}$)             &  0.149(21)   &  0.149(18)  & 0.1477(38)  & 0.1478(53)  \\
      \hline
   \end{tabular}
   \caption{Fits of the conic cut propagator data to equation (\ref{fit_uv}) and renormalization constants as function of the
                 the lattice volume. $q_{min}$ and $q_{max}$ stand for the lowest and highest momenta, respectively,
                 used in the fit to get $Z_R$.}
   \label{tab_Z}
\end{table}

All the simulations were performed on an hypercubic lattice which breaks rotational invariance.
In order to reduce lattice spacing effects, for each lattice and for momenta $q > 1$ GeV the conic cut
\cite{Leinweber99} was applied. For momenta below 1 GeV, all the data points were considered. 
In this way, we hope to have a good description of the infrared region.
The renormalization procedure, as described above, was performed separately for each lattice propagator. 
In all cases the fit to equation (\ref{fit_uv}) was smooth and the corresponding $\chi^2/d.o.f. \sim 1$.
The renormalized gluon propagator, after the cuts just described, is reported in figure \ref{fig_prop_all}. 
As can be seen in figure \ref{fig_prop_all}, finite volume effects are observed in our data.

In the following, we will consider two different analysis of the lattice data: (i)
a separate analysis for each lattice volume; (ii) the different lattices can be combined, after removing finite volume effects, 
to produce a propagator with a larger density of points in the momenta axis. If the different volumes allow for
an evaluation of the finite volume effects, the combined data, with its larger density of points on the $q$-axis, 
will reduce the statistical error on the final result.

Let us elaborate more on the combined data set.
From figure \ref{fig_prop_all} it is clear that the lattice propagator differ by more than one standard deviation
as a function of the physical volume. This effect is clearly seen in the infrared region.
To reduce the finite volume effects, we take the propagator from the largest volume considered here,  
corresponding to an $80^4$ lattice, as a reference and compare all the remain propagators to it.
To reduce the finite volume effects, the infrared data of the smaller lattices was removed if the propagator was not
compatible, within one standard deviation, with the $80^4$ propagator.
Due to the infrared cut, from the $64^4$ propagator only data with $q \ge 425$ MeV was considered, 
from $48^4$ only $q \ge 671$ MeV data was considered and from $32^4$ only data with $q \ge 848$ MeV was included.

The cuts just described do not remove all lattice spacing effects. Indeed, even after performing all the cuts one
can observe data points where, for the same $q^2$ the propagator differ by more than one standard deviation.
The difference is due to the violation of rotational invariance.
Therefore, to minimize such type of effects, for each lattice volume and for the same $q^2$ coming from 
different $q_\mu$, if the different estimates of the propagator don't agree within one standard deviation, 
one of the points is excluded. 
For example, for the $80^4$ lattice for momentum $q = 457$ MeV there are two estimates for the gluon propagator, 
$D(q^2) = 6.563(95)$ GeV$^{-2}$ associated with the lattice direction $n = (3, 0, 0, 0)$
and $D(q^2) =  6.416(71)$ GeV$^{-2}$ associated with the lattice direction $n = (2, 2, 1, 0)$, 
coming from different types of momenta.
The first value is clearly above all the data points and it was not considered in the data sets.
In this way, the surviving points will produce a unique curve for $D(q^2)$.

\begin{figure}[t]
   \centering
   \includegraphics[scale=0.35]{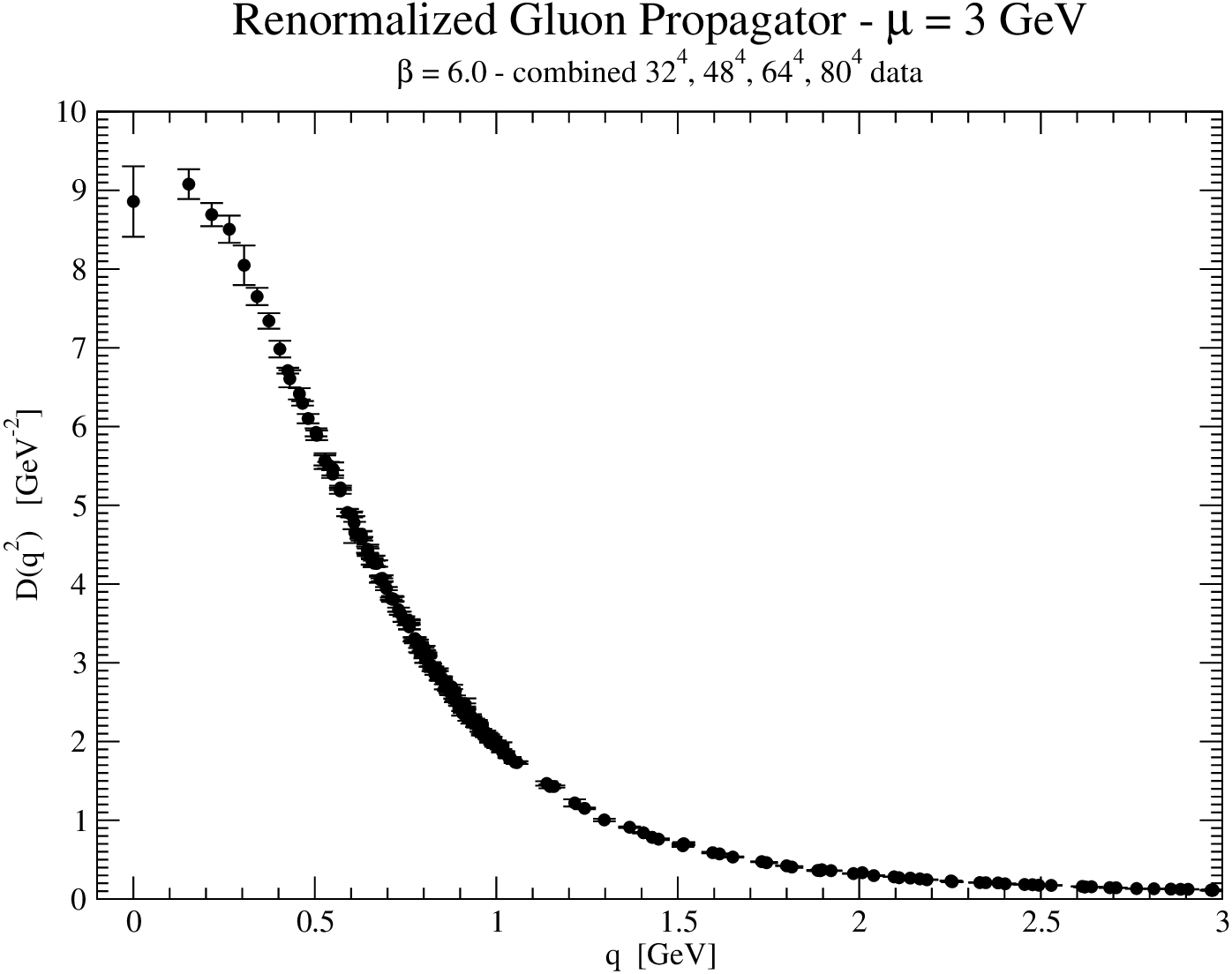} % requires the graphicx package
   \caption{Renormalized gluon propagator $D(q^2)$.}
   \label{fig_prop}
\end{figure}

\begin{figure}[t]
   \centering
   \includegraphics[scale=0.35]{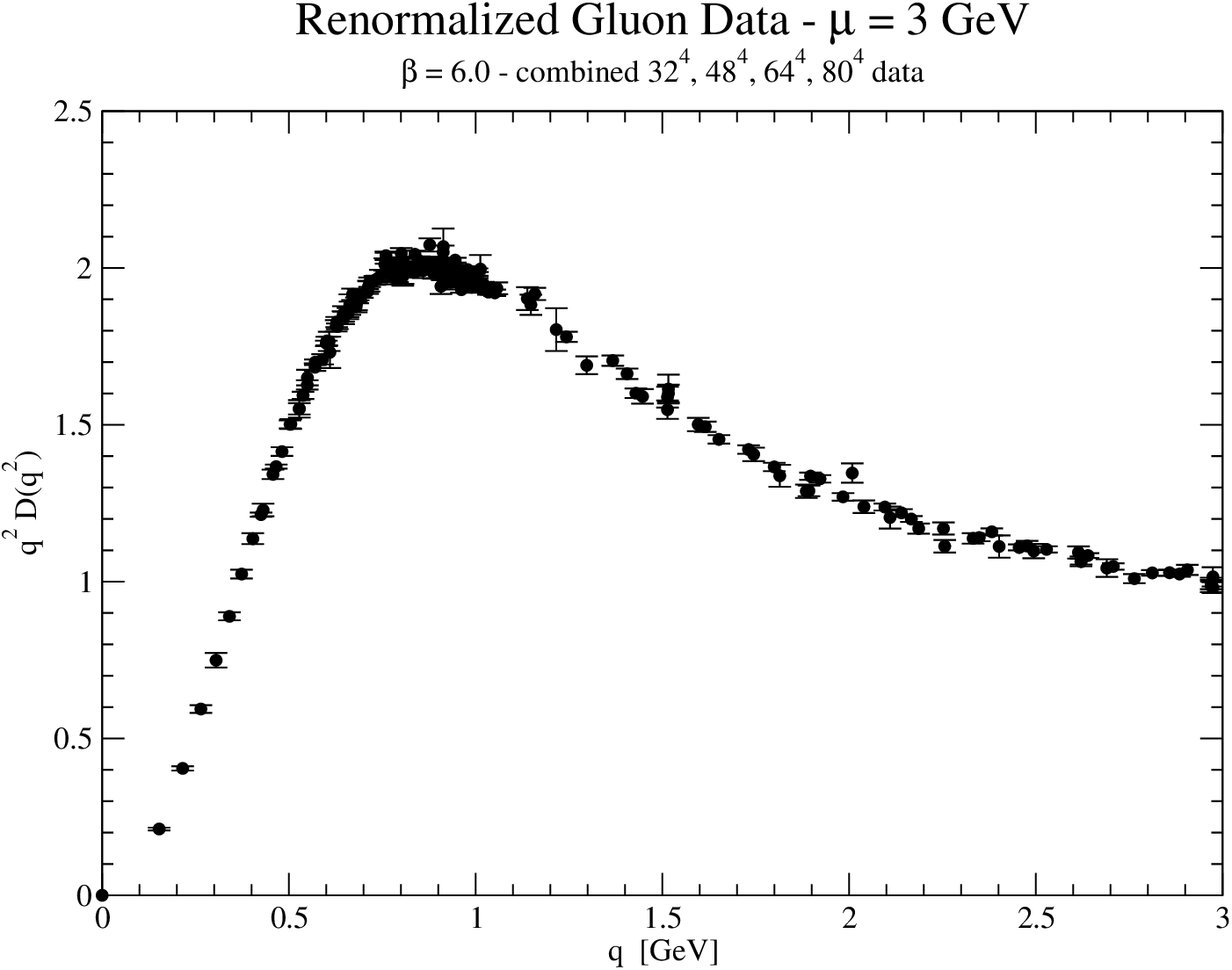} % requires the graphicx package
   \caption{Renormalized gluon data for $q^2 D(q^2)$.}
   \label{fig_dress}
\end{figure}

The combined lattice data for gluon propagator, after performing all the cuts,
is shown in figure \ref{fig_prop}. The corresponding $q^2 D(q^2)$ function is reported in figure \ref{fig_dress}.
Figures \ref{fig_prop} and \ref{fig_dress} suggest that both finite volume and finite lattice spacings have been removed from our
data.

The largest lattice volume included in our simulation is (8.1 fm)$^4$. Lattice
gluon propagators computed with larger volumes, but with a lattice spacing about twice the spacing considered in the present work, 
were reported in \cite{Bog_et_al_2009}. The two sets of data were compared in \cite{Dudal10}. 
We will not repeat this exercise but will resume the outcome of the comparison.
The propagator used in this work and those of \cite{Bog_et_al_2009} are essentially the same. Indeed, they
are compatible within one standard deviation for $q > 200$ MeV and show a small difference for smaller momenta.
For $q < 200$ MeV the propagator of \cite{Bog_et_al_2009} is about 10\% smaller than the data in figure 
\ref{fig_prop}. As seen in figure \ref{fig_prop_all}, our simulations have a limited access to momenta below 200 MeV and,
in this way, we expect that the impact of the finite volume effects on $M^2(q^2)$ computed from the combined data set is well
below 10\% factor. Furthermore, the separate analysis of the each volume will provide an estimate of the finite volume effects
on the running mass. 

Another source of systematics are Gribov copies, i.e. configurations which satisfy the Landau gauge condition
but are related by finite gauge transformations. 
This is a difficult and computational very demanding problem for the lattice practitioner.
However, the known SU(3) lattice simulations show that Gribov copies do not change significantly
the gluon propagator, i.e. that the effect due to the copies are, typically, within the statistical error; see, for example, \cite{Silva04}. Therefore,
in this work we will ignore possible effects due to the Gribov copies.

%====================================================================
%====================================================================
\section{The Gluon Mass \label{mass_gluao}}

Our goal is to obtain the gluon mass $M$. However, its computation, or the calculation of $M^2$, 
is not independent of the estimation of the dressing function $Z$. 
In the next sections we look at the gluon mass as given by the gluon propagator, 
i.e. by equation (\ref{Dmassive}), and explore different definitions.
First we consider constant mass ansatze to fit the propagator for the infrared and the ultraviolet momenta. 
However, the gluon propagator cannot be fitted over all momenta with a constant gluon mass. Then,
we proceed fitting the gluon propagator with a momentum dependent mass, called below running mass gluon,
and a running gluon dressing function.

%====================================================================
%====================================================================
\subsection{Constant Gluon Mass \label{section_const_mass}}

We start our analysis assuming a constant gluon mass, i.e. assuming that in equation (\ref{Dmassive}) 
$M$ and $Z$ are constants. Therefore, in this section we take the gluon propagator as being described by
\begin{equation}
   D(q^2) = \frac{Z}{q^2 + M^2} 
   \label{hard_mass}
\end{equation}
in a certain momentum range.

%====================================================================
%====================================================================
\subsubsection{Fitting of a Constant Infrared Gluon Mass \label{section_hadr_ir}}

\begin{figure}[t]
\vspace{0.3cm}
   \centering
   \includegraphics[scale=0.4]{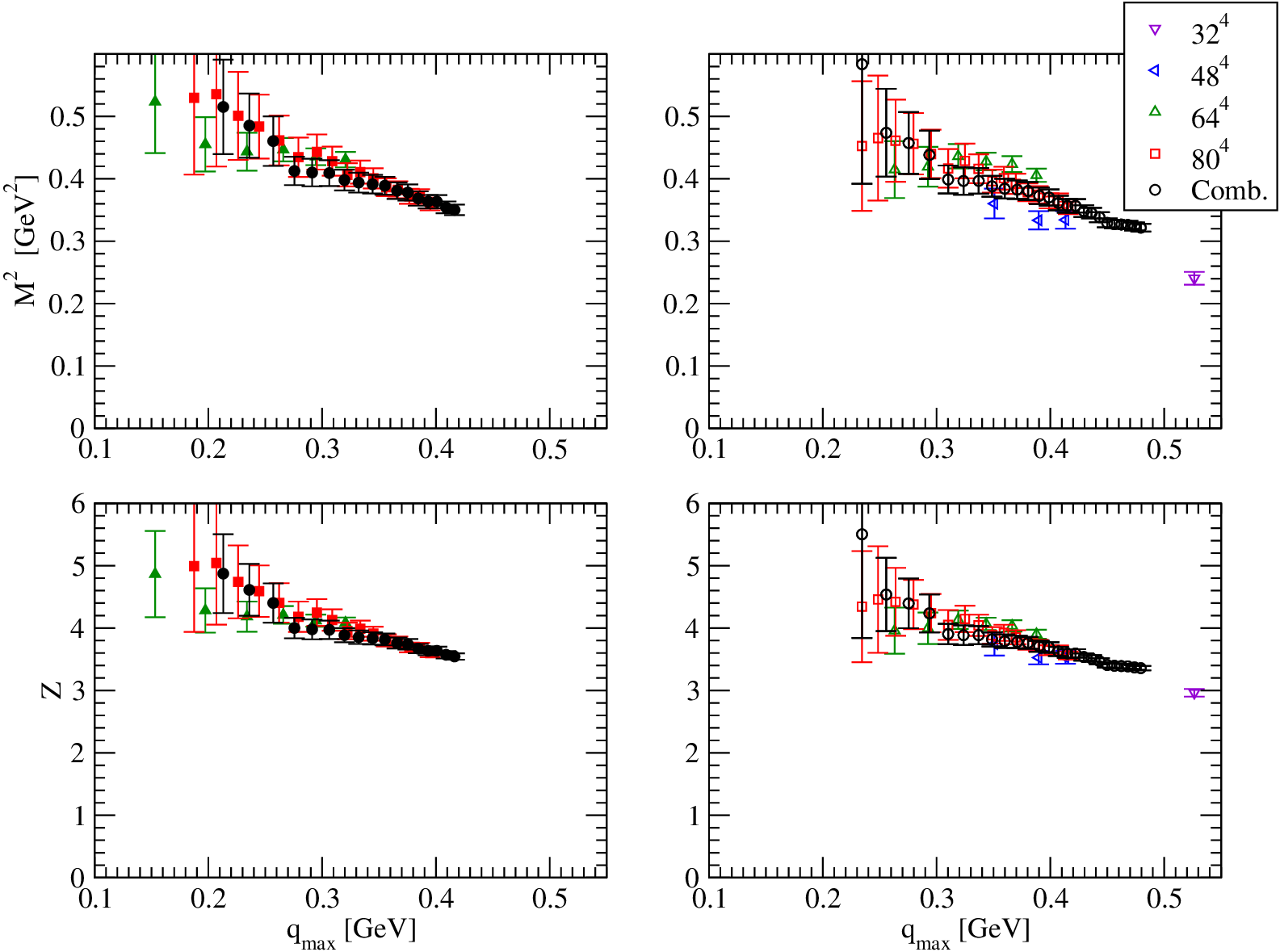} % requires the graphicx package
   \caption{Results of fitting the infrared lattice propagators to equation (\ref{hard_mass})
                 in the range $[0, q_{max}]$ and for $\chi^2/d.o.f.  \le 2.0$. In figure full symbols
                 refers to fits including the zero momentum propagator, while open symbols
                 mean that $D(0)$ was removed from the data sets.  \label{fig_hard_mass} }
\end{figure}

To check if the infrared gluon propagator can be described by such type of model, 
the lattice data was fitted to equation (\ref{hard_mass})
in the momentum range $[0, q_{max}]$. The results are plotted in figure \ref{fig_hard_mass}, where
full symbols mean that the zero momentum propagator was included in the fits, while open symbols
mean that $D(0)$ was removed from the data sets.

Figure \ref{fig_hard_mass} shows a $M^2$ and $Z$ that are, within one standard deviation, independent of the 
fitting range, i.e. of $q_{max}$. Furthermore, requiring a $\chi^2/d.o.f. < 2.0$ means that the infrared propagator
can be described by (\ref{hard_mass}) for momenta up to $q \sim 500$ MeV, if one ignores the lattice estimate for
$D(0)$, and up to $q \sim 430$ MeV, if the lattice $D(0)$ is included in the data set. More, only for volumes  of
$\sim$ (6.5 fm)$^4$ or larger, the infrared lattice propagator can be described by a massive type propagator
if $D(0)$ is taken into account in the data sets.

\begin{table}[t]
   \centering
   \begin{tabular}{l@{\hspace{0.7cm}}l@{\hspace{0.5cm}}l@{\hspace{0.5cm}}l@{\hspace{0.5cm}}l@{\hspace{0.5cm}}l} 
      \hline
 L                            &  32            &  48           & 64              & 80             & Comb.          \\
 \hline
 \multicolumn{6}{c}{Including $D(0)$} \\
 $q_{max}$ (MeV)  &   --            &  --             & 503            & 482           & 528               \\
 $Z$                        &   --            & --              & 4.082(86)   & 3.99(13)   & 3.760(87)      \\
 $M$ (MeV)            &   --            & --              & 656(10)      & 641(10)    & 617(11)            \\
 $\chi^2/d.of.$         &   --            & --             & 0.74            & 1.15         & 1.14               \\
 \hline
 \multicolumn{6}{c}{Excluding $D(0)$} \\
 $q_{max}$ (MeV)  &  660          &  508         & 503            & 505           & 550               \\
 $Z$                        & 2.963(62)  & 3.53(10)  & 4.035(89)   & 3.85(11)   & 3.594(68)      \\
 $M$ (MeV)            & 490(11)      & 578(12)   & 651(10)      & 626(13)   & 597(9)            \\
 $\chi^2/d.of.$        & 1.42            & 0.99        & 0.39            & 0.95        & 1.03               \\
      \hline
   \end{tabular}
   \caption{Results of fitting equation (\ref{hard_mass}). The values of $M$ and $Z$ reported
                 are for the fits whose $\chi^2/d.o.f$ is closer to unit for each data set.}
 \label{tab_fits_Mhard}
\end{table}

Values for $Z$, $M$ and $q_{max}$ for each data set are reported in table \ref{tab_fits_Mhard}. The table includes only the
results of the fits whose $\chi^2/d.of.$ is closer to unit. If one includes $D(0)$ in the data, $Z$ and $M$ seem to be essentially
independent of the volume. However, if one removes the zero momentum propagator from the data set, then one can observe
a slightly volume dependence with $M$ and $Z$ increasing with $V$. In both cases discussed, i.e. including of not
$D(0)$, $M$ and $Z$ computed from the combined data is just below the corresponding values obtained
from the two largest two volumes.

For the fits where $D(0)$ is excluded, a linear extrapolation of $M$ and $Z$ with $1/L$ towards the infinite volume
can be performed. Only after disregarding the $64^4$ results one is able to obtain excellent extrapolations.
Note that the numbers reported for the fits to the $64^4$ lattice propagator are off the main trend observed in the remaining fits.
The linear extrapolation to the infinite volume give
$Z = 4.49(10)$ with $\chi^2/d.o.f. = 0.38$ and 
$M = 723(11)$ MeV for a $\chi^2/d.o.f. = 0.70$. To these extrapolated values corresponds a 
\begin{equation}
     D(0) _{V = \infty} = 8.58(26) \qquad \mbox{GeV}^{-2},
     \label{D0_Vinfinito}
\end{equation}
in excellent agreement with the largest volume simulated here, $ D (0)_{V = (8.1 \mbox{ fm})^4} = 8.86(45) $ GeV$^{-2}$,
and with the infinite volume extrapolation computed in \cite{Dudal10}, $D (0)= 8.3(5)$ GeV$^{-2}$.
Further, the value (\ref{D0_Vinfinito}) agrees with the $D(0)$ computed using large volumes and reported in \cite{Dudal10}.

Our conclusion being that the fits show that the infrared lattice propagator is well described by a massive
type propagator from momenta up to $\sim  500$ MeV with an effective gluon mass  around $650 -  700$ MeV.
From table \ref{tab_fits_Mhard}, it follows that if $D(0)$ is included in the data set, then
a massive type propagator describes the lattice data up to moment $\sim 500$ MeV with $Z = 4.044(78)$ and
$M = 648(7)$ MeV given by the average values of the $64^4$ and $80^4$ figures; errors computed assuming gaussian
error propagation. On the other hand, if one excludes $D(0)$ from the data sets, again a massive type propagator describes
well the infrared lattice data up to $q \sim 500$ MeV with $Z = 4.49(10)$ and an effective gluon mass $M = 723(11)$ MeV.

%====================================================================
%====================================================================
\subsubsection{Fitting a Constant Ultraviolet Gluon Mass \label{section_hadr_uv}}

The same reasoning applied to infrared can be used to investigate the high momenta region. 
However, for the high momenta the fits to (\ref{hard_mass}) give a negative $M^2$, 
with $M^2$ depending strongly on the fitting range. 
We take this result as an indication that the ultraviolet is not described by such a propagator. 

Our discussion of the high momentum region is not in contradiction with the results of \cite{Leinweber99,Silva04},
where an ultraviolet gluon mass of $\sim 1 $ GeV was claimed. 
In \cite{Leinweber99,Silva04} an ultraviolet regulator was used and 
the lattice data surviving the conic cut fitted to
\begin{equation}
   D(q^2 ) = Z \, \frac{ \Big[ \frac{1}{2} \ln (q^2 + M^2) (q^{-2} + M^{-2}) \Big]^{-\gamma} }{q^2 + M^2}\, ,
   \label{UV_reg}
\end{equation}   
where $M$ is the gluon mass. 
Notice that the positive gluon mass in the numerator of eq. (\ref{UV_reg}) is equivalent to a negative mass in the
denominator of eq. (\ref{hard_mass}), and thus a negative ultraviolet mass, 
in the sense of eq. (\ref{hard_mass}), is not in contradistinction with perturbative QCD.

%====================================================================
%====================================================================
\subsection{Momentum Dependent Gluon Mass \label{section_run_m}}

The lattice gluon propagator can be described by a massive type propagator in the infrared region. For ultraviolet momenta, the
propagator follows closely the 1-loop perturbative prediction and, in this sense, one can claim that for the high momenta region
the effective gluon mass vanishes. Therefore, a way of recovering this two results is through the introduction of an effective
gluon mass which is a function of the gluon momenta.
So let us assume that $D(q^2)$ is given by equation (\ref{Dmassive}) and that $M(q^2)$ and the dressing function 
$Z(q^2)$ that are functions of the momentum. In the following we will refer to $M(q^2)$ and $Z(q^2)$ as the 
running mass and running dressing function, respectively.

Our first try to compute $M(q^2)$ and $Z(q^2)$ is to reproduce the analysis of the two previous sections. After
dividing the $q$-axis into small window momenta and, in each momenta window, fit the propagator assuming $Z$ 
and $M$ constant, 
the outcome is a decreasing $M^2(q^2)$ with $q^2$ and, for momenta above $\sim 1$ GeV, $M^2(q^2)$ becomes negative.
Given that the computation of $Z$ and $M^2$ are not independent, a negative mass squared is, certainly, due to a will conditioning
of at least one of the functions. Further, the problem of a negative mass squared can be cured, for example,
by changing the definition of $Z$.
This requires some model building and, instead, we will proceed fitting the lattice data to well known functional formulas.

Although, the outcome of this procedure to compute $M^2(q^2)$ and $Z(q^2)$ will not be reported here, we would like to call
the reader attention to some results found when analyzed the running mass and dressing functions obtained as described in the
previous paragraph. 

In what concerns $M^2(q^2)$, we find that in the infrared region it agrees well with the results reported in section 
\ref{section_hadr_ir}. Further, $M^2(q^2)$ follows a $q^2$ behavior in the infrared region and the lattice data can be
fitted assuming $M^2(q^2) \sim q^2 \ln q^2$ as suggest in \cite{Cornwall82}.

The running gluon dressing function $Z(q^2)$ decreases linearly from $q = 0$ up to 1 GeV. In the
ultraviolet region, $Z(q^2)$ agrees well with the perturbative QCD prediction. We observed that the gluon dressing
function is well described by the functional formula
\begin{equation}
   Z(q^2) = \frac{Z_0}{\left[ A + \ln(q^2 + m^2_0) \right]^\gamma} \, ,
   \label{z_fit_func0}
\end{equation}    
where $\gamma = 13/22$ is the anomalous gluon dimension and $m^2_0 \sim 1.6$ GeV$^2$. Note that
in equation (\ref{z_fit_func0}), $m^2_0$ plays a role similar to $M^2$ in (\ref{UV_reg}). 
In \cite{Leinweber99,Silva04}, the authors measured a $M = 1.0(1)$ GeV. The fits give essentially the same value,
i.e. an $m_0 \sim 1.3$ GeV.

\begin{figure}[t]
   \centering
   \includegraphics[scale=0.33]{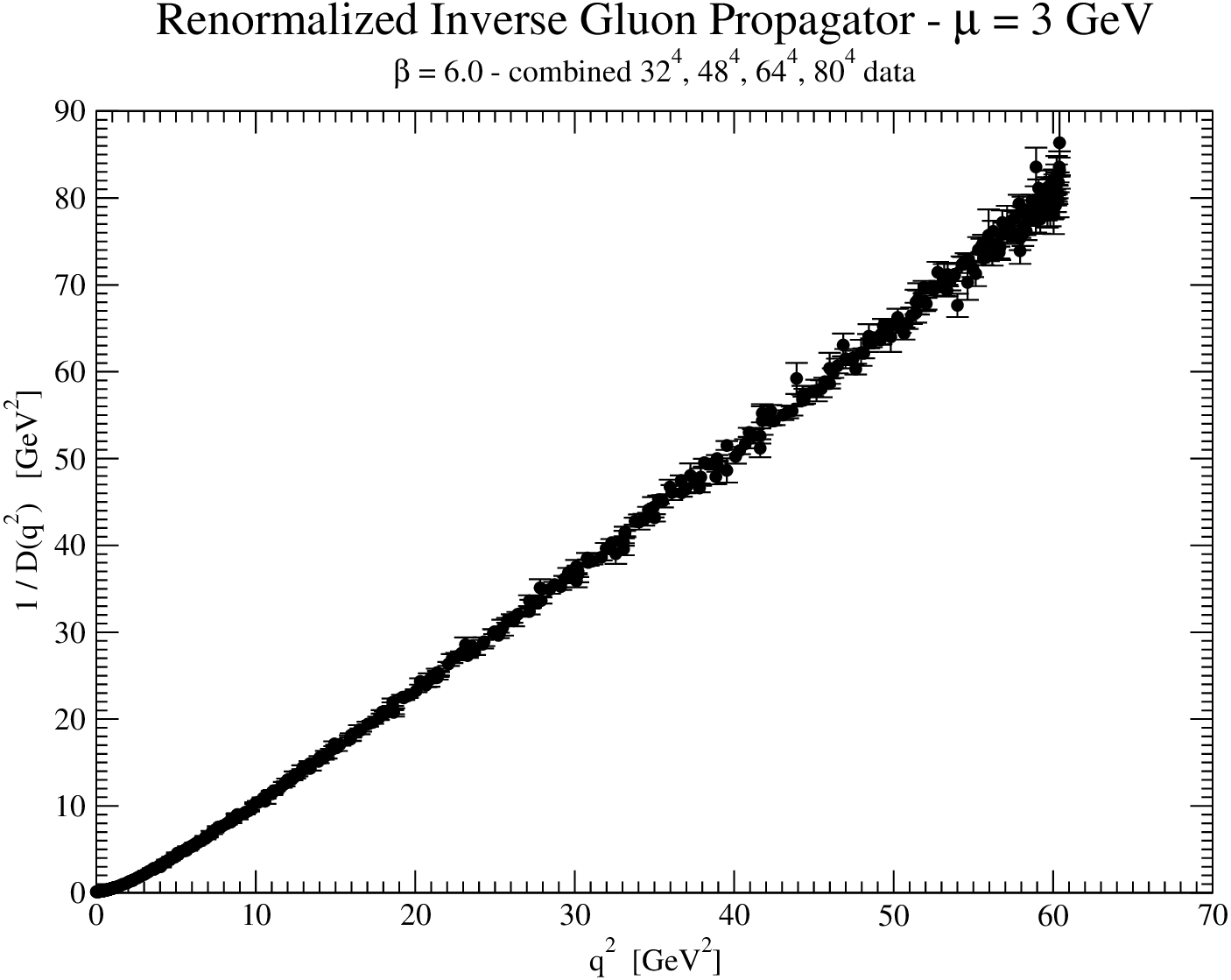} % requires the graphicx package
   \caption{Inverse gluon propagator for the combined data set. Note the "almost" linear behaviour with $q^2$ of $1/D(q^2)$.}
   \label{fig_inv_prop}
\end{figure}

In figure \ref{fig_inv_prop} the inverse of the propagator is plotted, for the combined data set, against $q^2$. The figure shows that
$1/D(q^2)$ follows essentially the perturbative behavior at large momenta, i.e. $ 1/D(q^2)  \propto q^2$, 
and at smaller momenta the data points deviate from a pure quadratic behavior. Not surprisingly, the observed ultraviolet 
confirms that $D(q^2)$ is well described by equation (\ref{fit_uv}).

%====================================================================
%====================================================================
\subsubsection{$M^2(q^2)$ from a Perturbative Inspired $Z(q^2)$:}

\begin{table}[t]
   \centering
   \begin{tabular}{l@{\hspace{0.5cm}}l@{\hspace{0.7cm}}l@{\hspace{0.5cm}}l@{\hspace{0.5cm}}l@{\hspace{0.5cm}}r} 
      \hline
 L        & $q_{min}$ (GeV)        &  $z$            &  $\Lambda$  (GeV)        & $\chi^2 / d.o.f.$             & \# d.of.                       \\
 \hline
 32  & 3        &  1.852(68)    & 0.726(79)  & 0.96  & 11 \\
       & 3.5     &  1.78(11)      &  0.83(14)   & 1.11  &   9  \\
       & 4        &  1.54(20)      &  1.22(39)   & 1.00  &  6   \\
 \hline
 48  & 3        &  1.830(75)    & 0.774(92)  & 1.21  & 15 \\
       & 3.5     &  2.00(12)      &  0.58(12)   & 0.78  &  11  \\
       & 4        &  2.16(29)      &  0.43(23)   & 1.01  &    8  \\
 \hline
 64  & 3        &  1.728(51)    & 0.879(71)  & 1.10  & 20 \\
       & 3.5     &  1.886(88)    &  0.67(10)   & 0.78  &  15  \\
       & 4        &  1.89(19)      &  0.66(22)   & 0.98  &   11  \\
 \hline
 80  & 3        &  1.761(86)    & 0.83(11)    & 1.27  & 25 \\
       & 3.5     &  2.05(18)      &  0.50(16)   & 1.04  & 19  \\
       & 4        &  2.04(28)      &  0.51(26)   & 0.96  &   14  \\
 \hline
 Comb  & 3        &  1.783(37)    &  0.812(47)   & 1.33  & 68 \\
            & 3.5     &  1.904(63)     &  0.661(72)   & 1.13  & 54  \\
            & 4        &  1.92(12)       &  0.64(14)     & 1.16  & 39  \\
      \hline
   \end{tabular}
   \caption{Results of fitting equation (\ref{extended_pert_D}) for $q \in [q _{min}, q_{max}]$ for $q_{max} = 5$ GeV.}
    \label{tab_extended_pert}
\end{table}

An effective gluon mass can be defined extending the perturbative behavior, observed in the high momentum region, 
towards the infrared region. Indeed, if one assumes that
\begin{equation}
   Z(q^2) = z    \left[  \log\left(1 + \frac{q^2}{\Lambda^2}\right) \right]^{ - \gamma } 
\end{equation}
is valid over all momentum range, which ensures that the gluon dressing function is finite everywhere, 
$z$ and $\Lambda$ can be measured fitting the ultraviolet lattice gluon data to
\begin{equation}
   D(q^2) = \frac{ z    \left[  \log\left(1 + \frac{q^2}{\Lambda^2}\right) \right]^{ - \gamma } }{q^2} \, ,
   \label{extended_pert_D}
\end{equation}
in a given range of momenta $[ q_{min}, q_{max} ]$. The running gluon mass is then defined by taking the lattice propagator
as
\begin{equation}
   D(q^2) = \frac{ z    \left[  \log\left(1 + \frac{q^2}{\Lambda^2}\right) \right]^{ - \gamma } }{q^2 + M^2(q^2)}
\end{equation}
for all $q$. With this definition, called below $M_{pert}(q^2)$, one ensures that the usual perturbative propagator is
recovered for high momenta or, equivalently, that $M_{pert}(q^2) \rightarrow 0$ for sufficently high $q$. As described below,
the gluon becomes massless for $q \sim 1$ GeV, which a surprisingly low mass scale. Of course, the exact momentum
scale at which the gluon becomes massless is connected with the parameterization of the gluon dressing function
$Z(q^2)$.

The results of fitting the lattice data to equation (\ref{extended_pert_D}), for $q \in [q _{min}, q_{max}]$ and
$q_{max} = 5$ GeV, are summarized in table \ref{tab_extended_pert}. The reason to exclude momenta above 5 GeV
is to avoid remaining lattice artifacts which where not removed by the renormalization procedure. In what concerns
$q_{min}$, it should belong to a region where the perturbative behavior is recovered. The renormalization procedure
described previously, see section \ref{definicoes} and table \ref{tab_Z}, 
suggests that $q_{min} = 3$ GeV belongs already to
the perturbative region. The two remaining $q_{min}$ values, i.e. 3.5 GeV and 4 GeV, where used to check for the 
stability of the fits. 

The results reported in table \ref{tab_extended_pert} show that, for all the data sets, equation (\ref{extended_pert_D})
describes well the lattice propagators in all momenta windows considered. 
Further, within each data set, $z$ and $\Lambda$ computed for the different
$q_{min}$ are compatible within one standard deviations. There are a few exceptions, where the $z$ and/or $\Lambda$
become compatible with the figures of the fits having the larger number of $d.o.f.$ within less than two standard deviations,
namely the fits to the $32^4$ data with the smallest momentum range, 
the $\Lambda$ from fits to the $48^4$ propagator with the smallest momentum range,
the fits to $64^4$ data with $q_{min} = 3.5$ GeV,
the $\Lambda$ from fits to the $80^4$ propagator with $q_{min} = 3.5$ GeV,
the fits to the combined data set with $q_{min} = 3.5$ GeV.
Moreover, comparing the fits with the larger number of degrees of freedom, it follows that $z$ and $\Lambda$, 
for all data sets are, compatible within one standard deviation.
Given the good stability of the fits, in the following, we will the results from the fits with the largest number of degrees of
freedom, i.e. for $q_{min} = 3$ GeV, to compute $M^2(q^2)$.

\begin{figure}[t]
\vspace{0.7cm}
   \centering
   \includegraphics[scale=0.4]{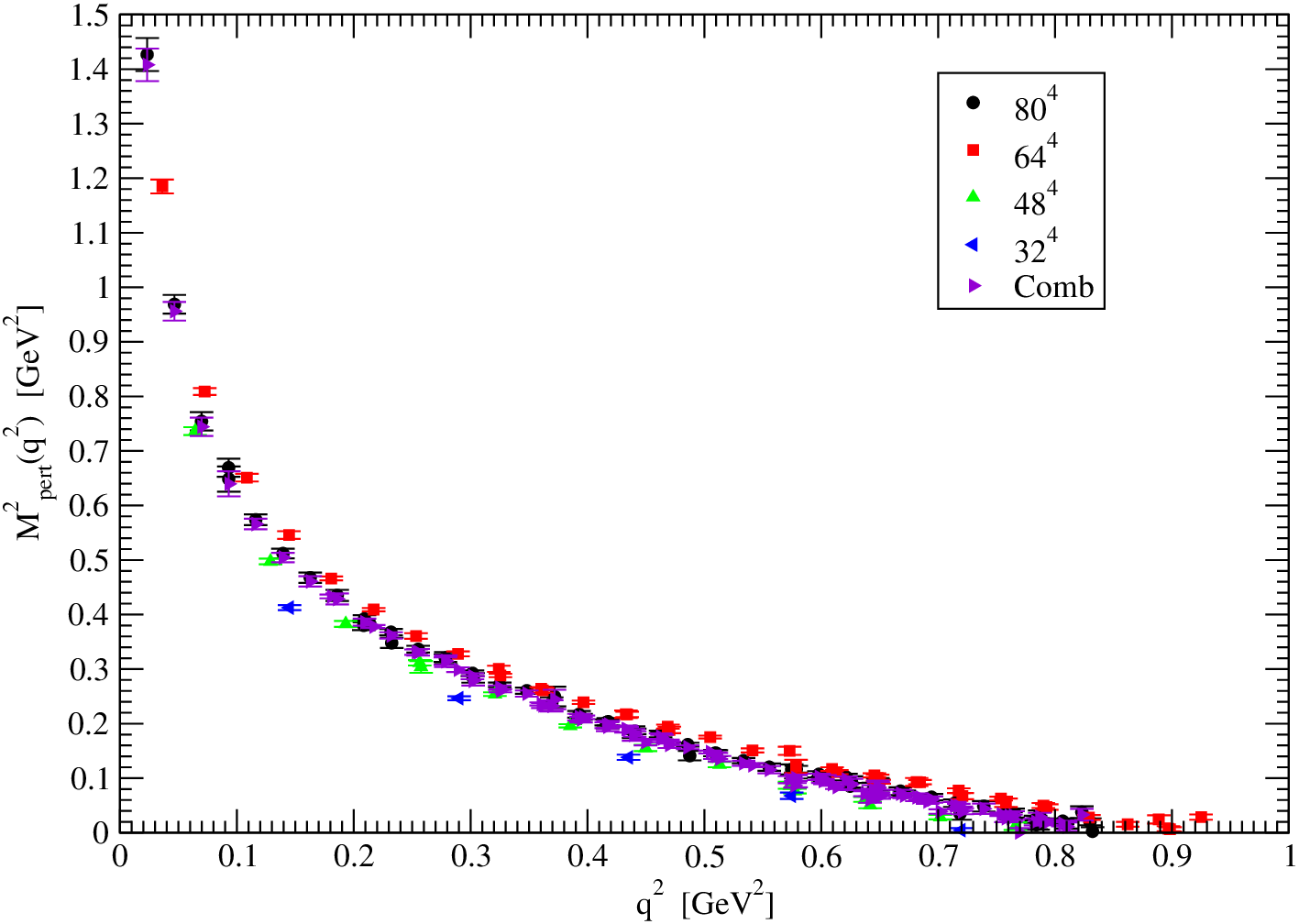} % requires the graphicx package
   \caption{$M^2_{pert} (q^2)$ for the various data sets. Only statistical errors from the propagators are take into account.
                 Statistical errors were computed assuming gaussian error propagation.}
   \label{fig_M2_ext_pert}
\end{figure}

Figure \ref{fig_M2_ext_pert} shows $M_{pert} (q^2)$ for all data sets.
The errors on $M^2_{pert} (q^2)$ include only the contribution form the lattice propagator and were computed
assuming gaussian error propagation. $M_{pert} (q^2)$ starts at value $\sim 1.2$ GeV for $q = 153$ MeV 
and decreases when $q^2$ increases. 
For all the data sets and all fits reported in table \ref{tab_extended_pert}, $M_{pert}(q^2)$ vanishes for momenta 
around 1 GeV. For the last point included in figure \ref{fig_M2_ext_pert}, the error bar is of the same order of
magnitude as the central value. For momenta above 1 GeV, not shown in figure \ref{fig_M2_ext_pert}, 
$M^2_{pert} (q^2)$ starts to fluctuate and the error bars increase.

The data for the different lattices have differences larger than one standard deviation, which is an indication of
finite volume effects. Certainly, the inclusion of the errors due to the parameterization of the gluon dressing function
used in the computation of $M^2_{pert} (q^2)$ improve the agreement between the different data sets.
Note, however, that the figure \ref{fig_M2_ext_pert} shows no clear systematic for the observed differences. 
Indeed, if the $64^4$ data is above the $80^4$ data, the $48^4$ data is below the largest
lattice considered in the present work.

Our interpretation of the results summarized in figure \ref{fig_M2_ext_pert} being that the gluon can be viewed
as having an effective mass for momenta below 1 GeV. For momenta above 1 GeV, the gluon propagator
follows closely the perturbative QCD prediction and becomes a massless boson.
This behavior helps explaining why equation (\ref{fit_uv}) could be used to described the lattice propagator for
momenta as low as 1.5 GeV - see the renormalization procedure and results described in table \ref{tab_Z} of section
\ref{definicoes}.

%====================================================================
%====================================================================
\subsubsection{Lattice $M^2(q^2)$ and Dyson-Schwinger Results:} 

As discussed in \cite{Cornwall82}, the nonperturbative solution of the QCD Dyson-Schwinger equations
allows for a gluon dynamical generated mass. Such a gluon mass was investigated in detail within the
so called decoupling solution of the DSE \cite{Aguilar03,Aguilar08,Aguilar08b,Aguilar08c,Aguilar10}. Typically,
the DSE are solved numerically and the numerical solution is fitted to a given functional form for the full momentum
range. By an appropriate choice of the fitting function, one can identify both
the running gluon dressing function and mass. Here, we would like to check if the same functional forms are able to
described the lattice propagator. Therefore, in this section, we will assume that the gluon propagator is given
by the (\ref{Dmassive}), with
\begin{equation}
  Z(q^2) = \frac{z_0 }{ \left[ \log \frac{q^2 \, + \, r \, m^2_0}{\Lambda}  \right]^\gamma } \, ,
\end{equation}  
where $\gamma$ is the anomalous gluon dimension, $m_0$ a mass scale and $r$ a numerical factor.

For the effective gluon mass, we will consider two different functional forms. A more sophisticated expression,
which incorporates the perturbative QCD ultraviolet prediction at high momenta
\cite{Aguilar08,Aguilar08b,Aguilar08c,Aguilar10},
\begin{equation}
  M^2_{soph}(q^2)  =  m^2(q^2,m^2_0) \,
                                         \left[ \frac{ \log \left( \frac{q^2 + f(q^2, m^2_0)}{\Lambda^2} \right)    }
                                                          {  \log \left( \frac{f(0, m^2_0)}{\Lambda^2} \right) }\right]^{- 3/5}  \, ,
\label{M_DSE_SOF}                                                          
\end{equation}  
with 
\begin{equation}
f(x,m^2_0) = \rho_1 \, m^2_0 \, + \, \rho_2 \, m^2(x,m^2_0) \qquad \mbox{ and } \qquad
m^2(x) = \frac{m^4_0}{x + m^2_0} \, .
\end{equation}
In the above expressions $\rho_1 = 1/2$ and $\rho_2 = 5/2$. 
Besides expression (\ref{M_DSE_SOF}), we will also consider the expression 
\begin{equation}
  M^2_{simp} (q^2) = \frac{m^4_0}{ q^2 + m^2_0} 
\label{M_DSE_SIMP}  
\end{equation}
for the effective gluon mass also used to described the numerical decoupling solution of the DSE.

\begin{table}[t]
   \centering
   \begin{tabular}{l@{\hspace{0.5cm}}l@{\hspace{0.7cm}}l@{\hspace{0.5cm}}l@{\hspace{0.5cm}}l@{\hspace{0.5cm}}l@{\hspace{0.5cm}}l} 
      \hline
 L        & $q_{max}$ (GeV)        &  $z_0$            &  $m_0$  (GeV)   & $\Lambda$ (GeV) & $r$ & $\chi^2 / d.o.f.$  \\
 \hline
 32  & 2.46        &  1.019(64)    & 0.756(13)  & 2.13(16)  & 8(1)    & 1.58 \\
 48  & 2.14        &  0.90(16)      & 0.751(43)  & 2.47(51)  & 11(6)  & 1.90 \\
 64  & 2.53        &  0.940(68)    & 0.711(17)  & 2.37(20)  & 12(3)  & 1.88 \\
 80  & 4.13        &  1.189(20)    & 0.706(16)  & 1.842(39)  & 7.49(59) & 1.74  \\
\hline
%   \end{tabular}
%   \caption{Fits of the lattice data to equation (\ref{M_DSE_SOF}). See text for details.}
%    \label{tab_DSE_SOF}
%\end{table}
%
%\begin{table}[t]
%   \centering
%   \begin{tabular}{l@{\hspace{0.5cm}}l@{\hspace{0.7cm}}l@{\hspace{0.5cm}}l@{\hspace{0.5cm}}l@{\hspace{0.5cm}}l@{\hspace{0.5cm}}l} 
%      \hline
% L        & $q_{max}$ (GeV)        &  $z_0$            &  $m_0$  (GeV)   & $\Lambda$ (GeV) & $r$ & $\chi^2 / d.o.f.$  \\
 \hline
 32  & 2.46        &  1.019(64)    & 0.756(13)  & 2.13(16)  & 8(1)    & 1.58 \\
 48  & 2.14        &  0.90(16)      & 0.751(43)  & 2.47(50)  & 11(6)  & 1.90 \\
 64  & 2.53        &  0.940(69)    & 0.711(17)  & 2.37(20)  & 12(3)  & 1.88 \\
 80  & 4.13        &  1.189(20)    & 0.706(16)  & 1.842(39)  & 7.49(59) & 1.74 \\
\hline
   \end{tabular}
   \caption{Fits of the lattice data using equation (\ref{M_DSE_SOF}) (upper part of table)
                  and (\ref{M_DSE_SIMP}) (lower part of table) for mass definition. See text for details.}
    \label{tab_DSE}
\end{table}

For the functional forms (\ref{M_DSE_SOF})  and (\ref{M_DSE_SIMP}), the parameters are
computed fitting the lattice data via minimization of the $\chi^2/d.o.f.$ 
For both expressions, we were not able to fit all momenta range. This can be understood
looking at the propagator data in ultraviolet region. Indeed, for large momenta,
the statistical errors on the lattice propagator are quite small and the minimization 
process is constraint mainly by these points. Further, the renormalization process does
not eliminate all the artifacts associated the finite lattice spacing. Indeed, a closer look at the
propagator show that for  $ q \sim 5$ GeV and above, the propagators fluctuates slightly. 
Given the small statistical errors, this makes extremely difficult to fit the data unless the 
fluctuations are removed as, for example, in \cite{soto07,Becirevic99,Becirevic00}. Instead
of smoothing our lattice data, we fitted (\ref{M_DSE_SOF})  and (\ref{M_DSE_SIMP}) from
$q = 0$ up to $q_{max}$, where $q_{max}$ was defined as the maximum momentum window
where $\chi^2/d.o.f. < 2$.

\begin{figure}[t]
\vspace{0.7cm}
   \centering
   \includegraphics[scale=0.4]{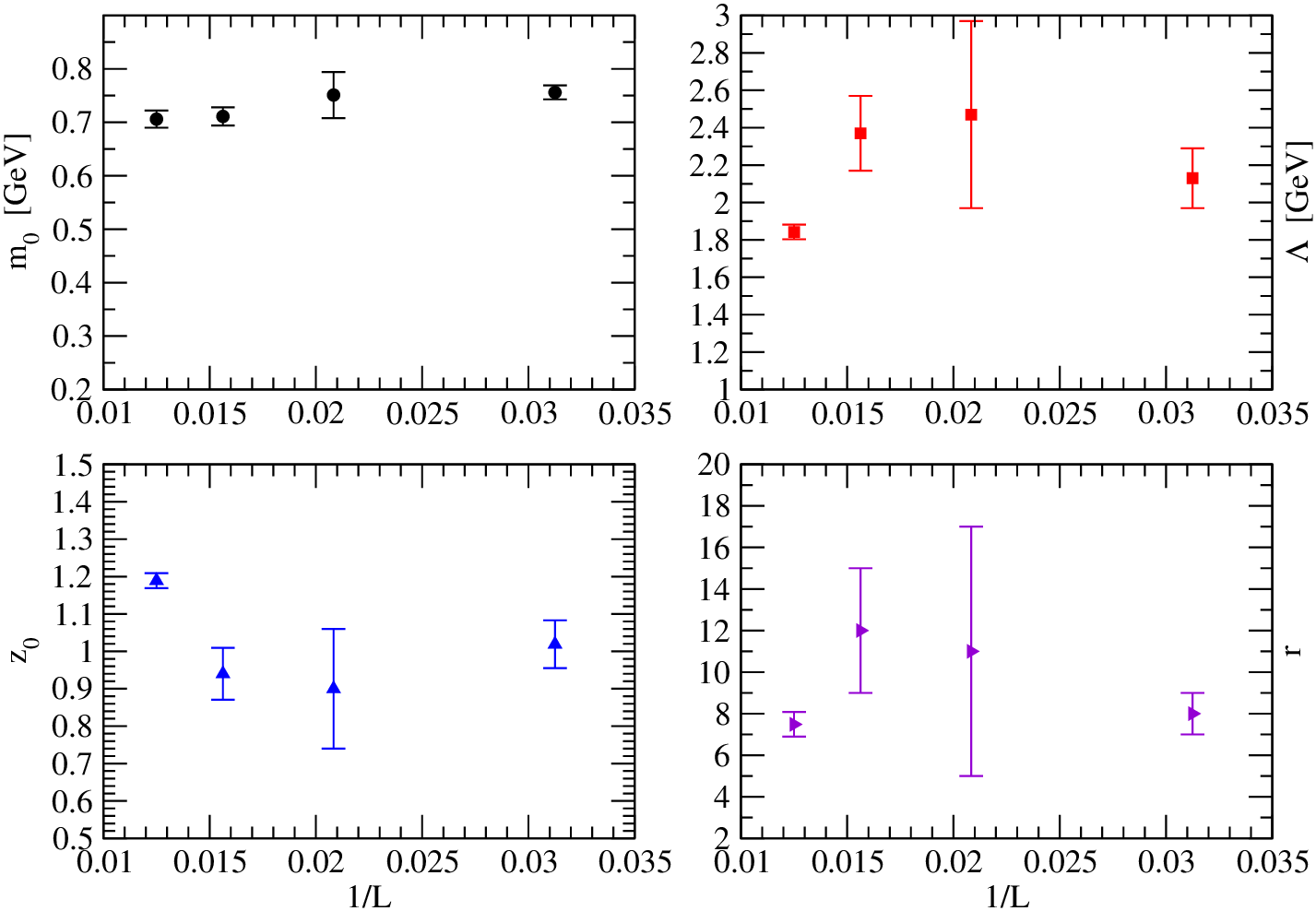} % requires the graphicx package
   \caption{Volume dependence of the parameters defined in (\ref{M_DSE_SOF}) and (\ref{M_DSE_SIMP}).}
   \label{fig_DSE_fits}
\end{figure}

The fits are reported in table \ref{tab_DSE}. Our first comment being that the fits are not able to distinguish between the two
functional forms (\ref{M_DSE_SOF}) and (\ref{M_DSE_SIMP}). The two functional forms describe quite well
the lattice propagator in the nonperturbative region, i.e. from zero momentum up to $\sim 4.2$ GeV. Furthermore,
the region which they are able to describe increases with the lattice volume. The exception being the smallest lattice, where
the good fit for a relatively large $q_{max}$ is related to the smaller number of $d.o.f.$, which makes the fit easier to perform. 

In the infrared region, (\ref{M_DSE_SOF}) and (\ref{M_DSE_SIMP}) reduce to a massive type propagator whose mass
is given by the mass parameter $m_0$. Looking at the infrared mass computed in section
\ref{section_hadr_ir}, see table \ref{tab_fits_Mhard}, it follows that the hard infrared effective gluon mass is slightly smaller
than $m_0$. However, $m_0$ is, within one standard deviation, compatible with the extrapolated infrared mass of 723(11) MeV.
Looking at $m_0$ obtained from the DSE equations, it turns out that their $m_0$ 
is slightly smaller than the numbers reported here. For example, in \cite{Aguilar10} the authors get for 
$m_0 = 612$ MeV. Note, however, that our definition does not match exactly their definition.

The volume dependence of the various parameters can be seen in figure \ref{fig_DSE_fits}. $m_0$ is well described by
a linear function of $1/L$, which can be used to extrapolate $m_0$ to infinite volume. The extrapolation to $V \rightarrow \infty$
gives $m_0 = 671(9)$ MeV, for a $\chi^2/d.o.f. = 0.16$. The extrapolated value is closer to the estimated
infrared hard mass measured directly from the lattice data measured in section \ref{section_hadr_ir}
and to the DSE estimate for $m_0$ reported in \cite{Aguilar10}.
For the remaining parameters, it is not clear which functional form should be used to extrapolate to infinite volume and,
in the following, we quote the value computed from the largest lattice. Note that all the points for $z_0$, $\Lambda$
and $r$ in figure \ref{fig_DSE_fits} are compatible within two standard deviations.

\begin{figure}[t]
\vspace{0.7cm}
   \centering
   \includegraphics[scale=0.4]{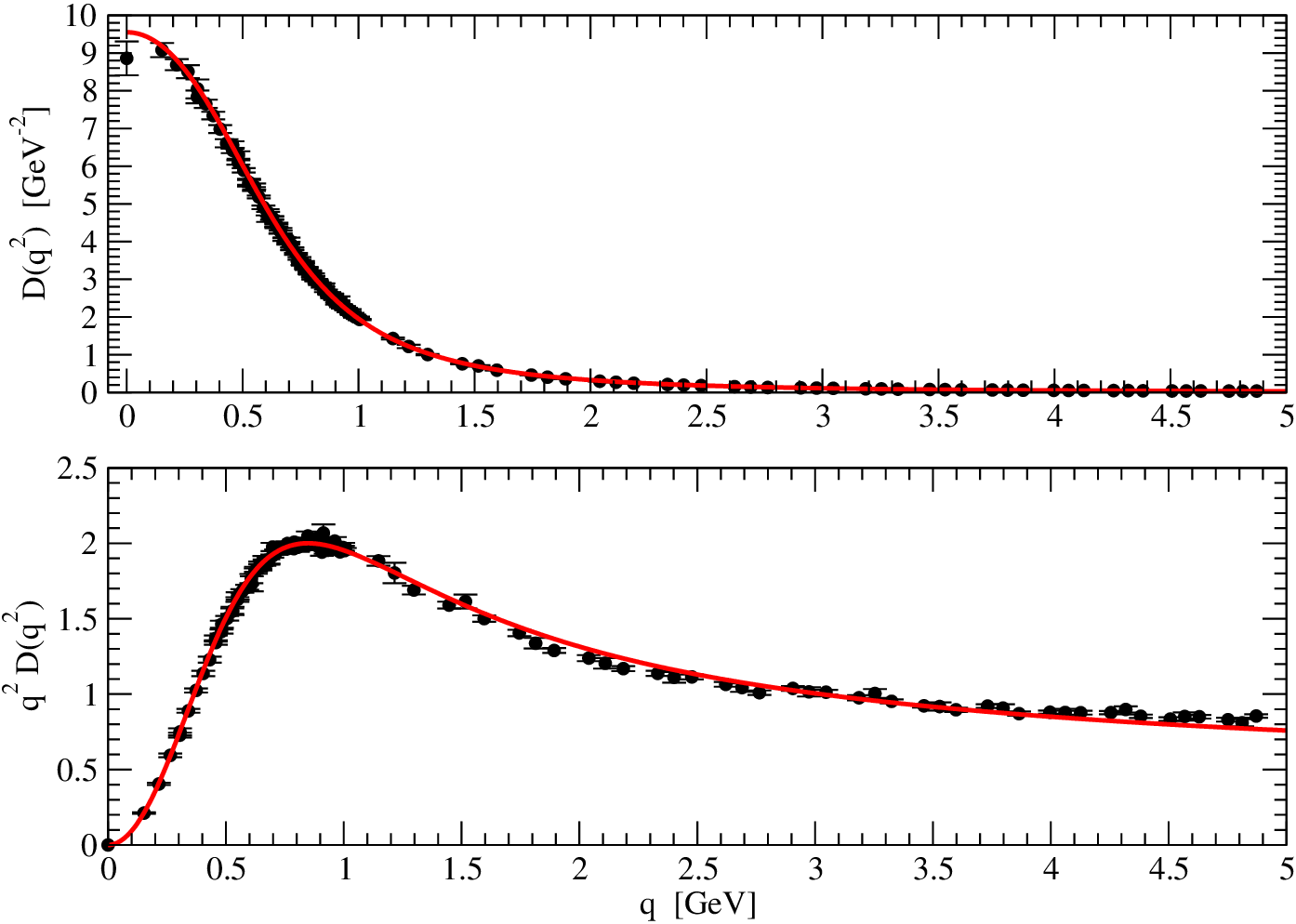} % requires the graphicx package
   \caption{Renormalized gluon propagator and $q^2 D(q^2)$ function from $80^4$ lattice data 
                 and corresponding using for the mass squared the definition (\ref{M_DSE_SIMP}).}
   \label{fig_DSE_80_fit}
\end{figure}

\begin{figure}[t]
\vspace{0.7cm}
   \centering
   \includegraphics[scale=0.4]{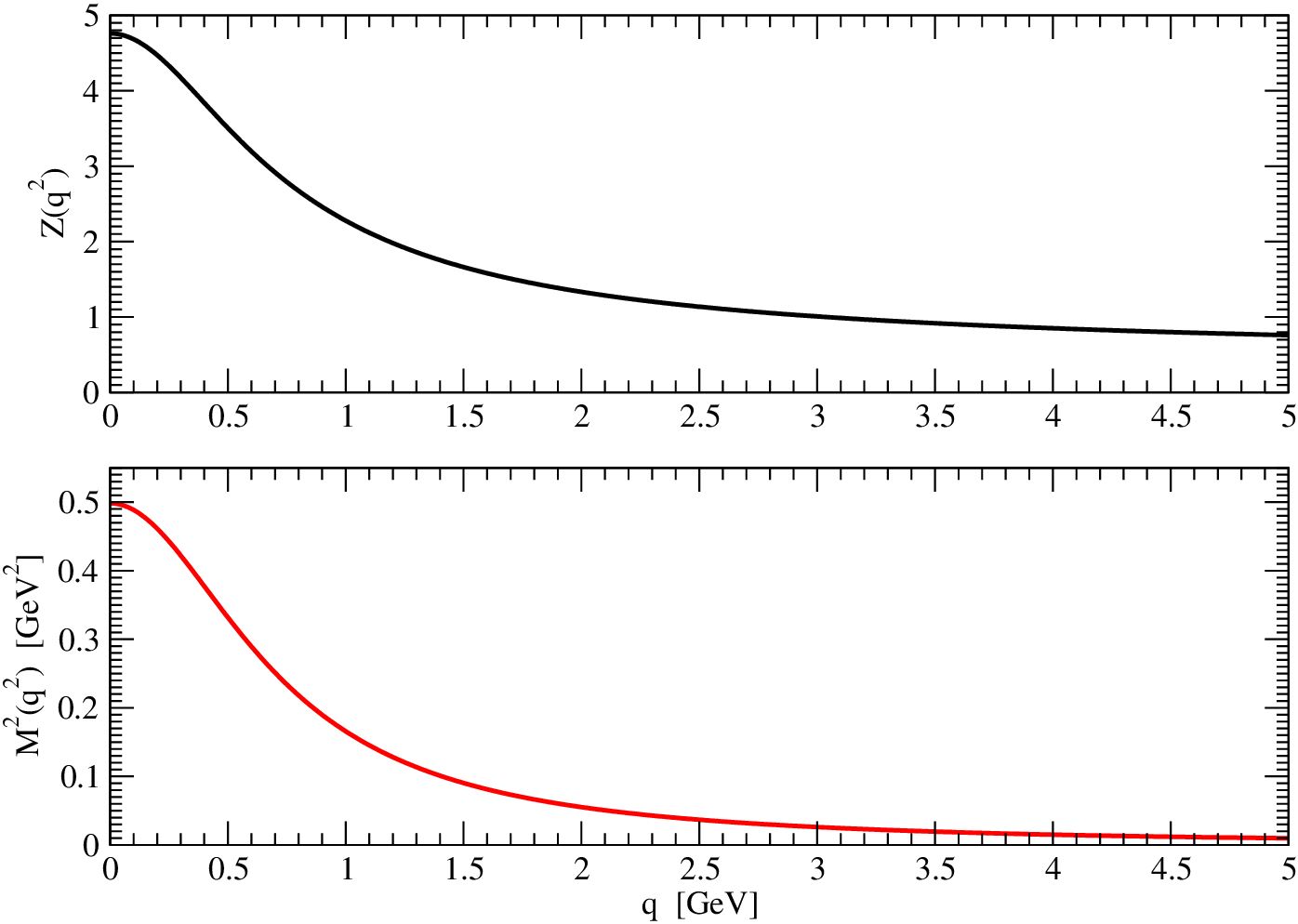} % requires the graphicx package
   \caption{Gluon dressing function and gluon running mass given by (\ref{M_DSE_SIMP}) from
                 the fits to the largest lattice volume.}
   \label{fig_DSE_Z_M2}
\end{figure}

From the previous analysis one can conclude that the functional forms used to described 
the gluon propagator from the decoupling solution of the Dyson-Schwinger also describe the lattice data. Furthermore,
they provide a definition for a dynamical generated gluon mass.

The renormalized gluon propagator and corresponding $q^2 D(q^2)$ function for the $80^4$ lattice,
together with the fits using (\ref{M_DSE_SIMP}) to parameterize the running gluon mass, are plotted in
figure \ref{fig_DSE_80_fit}. Figure \ref{fig_DSE_Z_M2} show the corresponding fitted dressing function
and gluon mass.

%====================================================================
%====================================================================
\section{Results and Discussion \label{fim}}

In this work we have investigated if the Landau gauge lattice gluon propagator can be described by a massive type 
propagator. For the
gluon mass itself, three different scenarios were considered: i) a constant infrared mass; ii) a constant ultraviolet mass; 
iii) a running mass in
association with a running gluon dressing function. The mass was measured from the momentum 
space gluon propagator given by equation (\ref{Dmassive}).

The interpretation of the infrared lattice gluon propagator as a massive propagator with constant $M$ and $Z$ 
was studied in section \ref{section_hadr_ir}. Our results show that the lattice data is compatible with such a picture
for momenta up to $\sim 500$ MeV, with an effective gluon mass around 650 - 700 MeV. If the zero momentum gluon
propagator is included in the data sets, then the measured $Z = 4.044(78)$ and $M = 648(7)$ MeV.
On the other hand, if one removes $D(0)$ from the analysis, $Z = 4.49(10)$ and the effective gluon becomes
slightly higher $M = 723(11)$ MeV.

The measured infrared hard mass reproduces the results reported in \cite{Oliveira09} for 
SU(3) simulations and is consistent with the quoted infrared mass value estimated for SU(2) in \cite{Bornyakov_2009}, 
where an $M$ of 0.69(3) GeV or 0.68(4) GeV, depending if one includes or not 
include $D(0)$ on the analysis, was claimed. Furthermore, the gluon masses claimed above agree
well with the estimate of the gluon mass from the gluon condensate $\langle A^2 \rangle$ 
obtained in \cite{arriola04}, where the value 625(33) MeV was obtained, and is well within the interval of values
estimated from phenomenology \cite{Field02}. 

The interpretation of the lattice gluon propagator as massive type propagator with a 
constant mass for the ultraviolet region
was checked in section \ref{section_hadr_uv}. 
It turns out that the lattice data cannot be fitted consistently by such a propagator. $M^2$ depends on the 
fitting range $[ q_{min}, q_{max}]$ and, for each $q_{max}$, $M^2$ is not constant. This means that, in the ultraviolet, 
the lattice gluon propagator does not behave as a massive bosonic propagator with a non-vanishing
constant mass.

The case of a running gluon mass $M^2(q^2)$ and a running dressing function $Z(q^2)$ 
was studied in section \ref{section_run_m}. Several definitions for $M^2(q^2)$ together with
$Z(q^2)$ where investigated. In general, it turns out that $M^2$ is a decreasing function of $q$.
Of course, the precise value for $M^2(q^2)$ depends on the chosen definition for $Z(q^2)$ 
and vice-versa. Here we considered two cases for the dressing function which allow for the computation
of the running mass: (i) a na\"{\i}ve extension of the perturbative dressing function towards the infrared, 
keeping $Z(q^2)$ always finite; (ii) the functional forms for $M^2(q^2)$ and $Z(q^2)$ used to fit 
the so-called decoupling type solution of the gluon-ghost Dyson-Schwinger equations.

Our first computation of the gluon running mass considered a parameterization of the
running dressing function which extended the perturbative $Z$ towards the infrared region,
while keeping $Z(q^2)$ finite for all momenta. The gluon mass defined in this way,
$M^2_{pert}(q^2)$ is shown in figure \ref{fig_M2_ext_pert}. $M(q^2)$ runs from $\sim 1.2$ GeV
for $q= 0$ down to zero for $q \sim 932$ MeV. We would like to call the reader attention, that if
instead of using the results reported in table \ref{tab_extended_pert} for the larger fitting range, we
used the results for the smallest fitting range, the values for $M_{pert}(q^2)$ would be reduced,
starting below 1 GeV for the smallest momentum. 
However, the qualitative behavior will be similar to that observed in figure
\ref{fig_M2_ext_pert}.

After investigating $M^2_{pert}(q^2)$, we have checked the compatibility between the 
functional forms used to described the decoupling type solution of the Dyson-Schwinger
equations for the gluon propagator and the lattice propagators. These functional forms
provide a parametrization of the gluon propagator and a definition for the running
gluon mass. Moreover, they take into account the perturbative QCD predictions for the ultraviolet regime.
The functional forms are able to describe quite well the lattice data over the entire non-perturbative
regime. The lattice data can be fitted by the expressions considered in the present work
from $q = 0$ up to $q_{max}$, with $q_{max}$ increasing with the lattice volume. 
For our largest volume $q_{max} = 4.13$ GeV. For momenta above $\sim 4.13$ GeV the remaining lattice spacing effects
prevent a fit to the data.

In what concerns the gluon mass, we found that it is well described by
\begin{equation}
   M^2(q^2)  = \frac{m^4_0}{q^2 + m^2_0} \, ,
\end{equation}   
with $m_0 = 671(9)$ MeV. Our estimation of $m_0$ agrees well with the estimation performed using the DSE and 
with is, within errors, compatible with the infrared hard mass when the $D(0)$ was included in the analysis.
 
From the above results it follows that the interpretation of the gluon propagator as a massive type propagator 
with a momentum dependent mass and dressing function fits, quite well, the lattice QCD data. 
The gluon mass is a decreasing function of $q^2$ and becomes massless in the ultraviolet region.
The nature of the gluon mass helps in the understanding of the remarkable mechanism of confinement, 
where the gluon mass may contribute to a Meissner-type effect in QCD. Moreover, given that one can 
define an effective gluon mass, it follows that the $\langle A^2 \rangle$ should be taken into account
when investigating the non-perturbative dynamics of QCD.

%====================================================================
%====================================================================
\ack

The authors acknowledge financial support from F.C.T. under project CERN/FP/83582/2008 and CERN/FP/109327/2009.
OO aknowledges FAPES for financial support.
The authors thank A. Aguilar for helpful discussions. The authors thank P. J. Silva for the help 
with the gauge fixing for the $32^4$ lattice.

%%%%%%%%%%%%%%%%%%%%%%%%%%%%%%%%%%%%%%%%%%%%%%%%%%%%%%%%%%%%%%%%%%%%%%%%%%%%%%%%%%%%%%%%%%%%%%%%%%%%%%%%%%%%%%%%%%%%%%%%

%%%%%%%%%%%%%%%%%%%%%%


\begin{thebibliography}{99}

\bibitem{FaPo67} Faddeev L P  and Popov V N 1967 \textit{Phys. Lett}  B \textbf{25} 29

\bibitem{Cornwall82} Cornwall J H  1982 \textit{Phys. Rev.} D \textbf{26} 1453

\bibitem{Forshaw99}
 Forshaw J R, Papavassiliou J and  Parrinello C 1999  \textit{Phys. Rev.} D \textbf{59} 074008

\bibitem{Field02}
 Field J H 2002 \textit{Phys. Rev.} D \textbf{66} 013013
 
 \bibitem{Oliveira09}
Oliveira O and SIlva P J 2009 Pos (\textbf{QCD-TNT09}) 33 (2009) [{\tt arXiv:0911.1643}].

\bibitem{Dudal04}
Dudal D, Verschelde H, Gracey J A, Lemes V E R, Saranday M S, SObreiro R F and
Sorella S P (2004) \textit{JEHP} \textbf{401} 44

\bibitem{Esole04}
Esole M and Freire F (2004) \textit{Phys. Rev.} D \textbf{69} 41701

\bibitem{Dudal10}
Dudal D, Oliveira O and Vandersickel N 2010 \textit{Phys. Rev.} D \textbf{81} 074505

\bibitem{Aguilar03} 
    Aguilar A C, Natale A A and  Rodrigues da Silva P S 2003  \textit{Phys. Rev. Lett.} \textbf{90}  152001

\bibitem{Aguilar08}    
   Aguilar A C and Papavassiliou J 2008 \textit{Phys. Rev.} D \textbf{77} 125022

\bibitem{Aguilar08b}   
   Aguilar A C, Binosi D and Papavassiliou J 2008 \textit{Phys. Rev.} D \textbf{78} 025010
   
\bibitem{Aguilar08c}   
 Binosi D and Papavassiliou J 2008 \textit{JHEP} \textbf{811} 63
 
 \bibitem{Aguilar10}
Aguilar A C and Papavassiliou J 2010 \textit{Phys. Rev.} D \textbf{81} 034003 [arXiv:0910.4142] 

\bibitem{Cornwall09}
 Cornwall J M 2009 \textit{Phys. Rev.} D \textbf{80} 096001



\bibitem{Sauli09}
Sauli V {\tt arXiv:0906.2818 [hep-ph]}

\bibitem{Fischer09}
Fischer C S, Maas A and Pawlowski J M 2009 \textit{Ann. Phys.} \textbf{324} 2408


\bibitem{Dudal11}
Dudal D, Guimar\~ares M S and Sorella S P, 2011 \textit{Phys. Rev. Lett.} \textbf{106}, 062003 [arXiv:1010.3638]

\bibitem{Dudal08}
Dudal D, Gracey J, Sorella S P, Vandersickel N and Verschelde H 2008 \textit{Phys. Rev.} D \textbf{78} 065047


\bibitem{Leinweber99}
Leinweber D B, Skullerud J I, Williams A G and  Parrinello C 1999 \textit{Phys. Rev.} D \textbf{60} 094507

\bibitem{Silva04}
Silva P J and Oliveira O 2004 \textit{Nucl. Phys.} B \textbf{690} 177

\bibitem{Bernard94}
Bernard C, Parrinello C and Soni A 1994 \textit{Phys. Rev.} D \textbf{49} 1585



\bibitem{Bali93}
Bali G S and Schiling K 1993 \textit{Phys. Rev.} D \textbf{47} 661

\bibitem{milc}
This work was in part based on the MILC collaboration's public lattice gauge theory code:
http://physics.indiana.edu/\~sg/milc.html.

\bibitem{Bog_et_al_2009}
Bogolubsky I L,  Ilgenfritz E-M, M\"uller-Preussker M and Sternbeck A 2009 \textit{Phys. Lett.} B \textbf{676} 69


\bibitem{soto07}
Soto F de and Roiesnel C 2007 \textit{JHEP} \textbf{709} 7

\bibitem{Becirevic99}
Becirevic D, Boucaud Ph, Leroy J P, Micheli J, P\`ene O, Rodr\'{\i}guez-Quintero J and Roiesnel C
1999 \textit{Phys. Rev.} D \textbf{60} 094509 [arXiv:hep-ph/9903364]

\bibitem{Becirevic00}
Becirevic D, Boucaud Ph, Leroy J P, Micheli J, P\`ene O, Rodr\'{\i}guez-Quintero J and C. Roiesnel 
2000 \textit{Phys. Rev.} D \textbf{61} 114508 [arXiv:hep-ph/9910204]





\bibitem{Bornyakov_2009}
Bornyakov V G, Mitrjushkin V K and M\"uller-Preussker M {\tt arXiv:0912.4475}

\bibitem{arriola04}
Arriola E R, Bowman P O and Broniowski W 2004 \textit{Phys. Rev.} D \textbf{70} 097505


\end{thebibliography}
\end{document}